\documentstyle[12pt]{article}
\newcommand{\be}{\begin{equation}}
\newcommand{\ee}{\end{equation}}

\newcommand{\bi}[1]{\bibitem{#1}}
\newcommand{\fr}[2]{\frac{#1}{#2}}

\newcommand{\gm}{\mbox{$\gamma_{\mu}$}}
\newcommand{\ph}{\mbox{$\hat{p}$}}
\newcommand{\qh}{\mbox{$\hat{q}$}}

\newcommand{\FD}{\mbox{$\tilde{F}$}}
\newcommand{\gf}{\mbox{$\gamma_{5}$}}

\newcommand{\hpl}[5]{ \put(#1,#2){\begin{picture}(80,20)
 \multiput(#3,0)(#4,0){#5}{\oval(#3,#3)[b]}
 \multiput(0,0)(#4,0){#5}{\oval(#3,#3)[t]}  \end{picture}}  }
\newcommand{\floop}[2]{ \put(#1,#2){\begin{picture}(60,60)(0,0)
\put(10,60){\oval(15,30)}
\hpl{3}{68}{1.5}{3}{5}
\hpl{-6}{52}{1.5}{3}{3}
\hpl{18}{52}{1.5}{3}{3}
\end{picture}}}

\begin{document}
\normalsize
\begin{flushright}{BINP 93--88 \\ October 1993}
\end{flushright}
\vspace{1.0cm}
\begin{center}{\Large \bf Magnetic quadrupole moment of $W$-boson in
Kobayashi-Maskawa model}\\
\vspace{1.0cm}
{\bf I.B. Khriplovich\footnote{khriplovich@inp.nsk.su}
and M.E. Pospelov\footnote{pospelov@inp.nsk.su}}\\

Budker Institute of Nuclear Physics, 630090 Novosibirsk
\vspace{4.0cm}
\end{center}

\begin{abstract}
Due to CP-invariance violation a vector particle can acquire T-
and P-odd electromagnetic moment, magnetic quadrupole one. The W-boson
magnetic quadrupole moment is calculated in the Kobayashi-Maskawa model.
This is the only known CP-odd moment arising in this model in two-loop
approximation.
\end{abstract}
\newpage

1. The Kobayashi-Maskawa (KM) model looks now as the most natural
description of CP-violation. It describes properly CP-odd phenomena
in the decays of neutral $K$-mesons and predicts extremely tiny CP-odd
effects in the flavour-conserving processes. Though its predictions for
the electric dipole moments (EDM) of elementary particles are far beyond
the present experimental facilities, the corresponding theoretical
investigations are of certain methodological interest.

To lowest, one-loop approximation in the weak interaction all CP-odd
flavour-conserving amplitudes in the KM model turn to zero trivially. The
point is that in this approximation those amplitudes depend only on the
moduli squared of elements of the KM matrix, so the result cannot contain
the CP-violating phase.

However, the EDMs of a quark and W-boson vanish also to next, two-loop
approximation as well\cite{Shab,KhP}. It has no transparent explanation,
though in Ref.\cite{ckl} an attempt was made to explain vanishing of the
W-boson EDM in two-loop approximation starting from general principles
only.

In the present work the W-boson magnetic quadrupole moment (MQM) is
calculated in the KM model. It arises already in two-loop approximation
which means in particular that the vanishing of EDM in this
approximation is a very specific, dynamic phenomenon only.

To avoid a possible misunderstanding we wish to mention here that the
interaction of the W-boson MQM with an external electromagnetic field,
being of second order in the field momentum $q$, does not induce for instance
the electron EDM since its interaction with an electromagnetic field is
of first order in $q$.

2. A general consideration ascending to Ref.\cite{DKO} demonstrates that a
particle of spin $I$ can have $2I$ P- and T-odd electromagnetic
moments. In particular, at $I=1$ there are two such moments. Besides
an electric dipole moment a particle of spin one can have a magnetic
quadrupole moment.

We define the MQM operator by analogy with the electric quadrupole one,
via the interaction with the corresponding field gradient:
\begin{eqnarray}
\hat{H}_{Q}=-\fr{1}{6}\hat{Q}_{ij}\nabla_iE_j, \nonumber\\
\hat{H}_{M}=-\fr{1}{6}\hat{M}_{ij}\nabla_iB_j.
\label{eq:def}
\end{eqnarray}
The tensor $\hat{M}_{ij}$ is expressed as usual via the irreducible
second-rank tensor constructed from the spin operator
$\hat{I}_i$:
\be
\hat{M}_{ij}={\cal M}\fr{3}{2I(2I-1)}[\hat{I}_i\hat{I}_j+\hat{I}_j\hat{I}_i
-\fr{2}{3}\delta_{ij}I(I+1)].
\ee
The expectation value ${\cal M}$ of the operator $\hat{M}_{zz}$
in the state with the maximum spin projection $I_z=I$ will be called
magnetic quadrupole moment.

Let us consider now the effective interaction of the W-boson with
magnetic field gradient. Since the spin operator acts as follows on the
polarization vector $\bf{e}$ of a vector particle at rest \cite{LL}:
\be
(\hat{I}_i\bf{e})_k =-i\epsilon_{ikl}e_l,
\ee
for such a particle the matrix element of the MQM interaction reduces to
\be
H=\fr{1}{4}{\cal M}e^*_i e_j(\nabla_i B_j+\nabla_j B_i).
\ee
The covariant form of this interaction is
\be
H=-\fr{1}{2}{\cal M}W_{\mu}^*W_{\nu}(\partial_{\mu}\FD_{\nu\alpha}+
\partial_{\nu}\FD_{\mu\alpha})k_{\alpha}.
\label{eq:MQM}
\ee
Here  $k$ is the 4-momentum of the
W-boson, $W_{\mu}=e_{\mu}/\sqrt{2k_0}$, $e_{\mu}=(e_0,{\bf e})$
is its covariant polarization vector,
$\partial_{\mu}=\partial/\partial x_{\mu}$,
\,$\FD_{\mu\nu}=\fr{1}{2}\epsilon_{\mu\nu\alpha\beta}F_{\alpha\beta}$.

3. We are going over now to the direct calculation of the W-boson MQM
in the standard model to two-loop approximation. The general structure of
the diagrams which could contribute to the effect in that approximation is
presented in Fig.1. The external photon line can be attached on to any
quark or W propagator. To first order in the external photon momentum
$q$, which corresponds to the EDM interaction with a constant external
field, the sum of those diagrams vanishes \cite{KhP}. Now however we are
interested in the terms of second order in $q$.

The CP-odd part of the loop flavour structure is:
\begin{eqnarray}
i\tilde{\delta}[d(c(b-s)t-t(b-s)c+t(b-s)u-u(b-s)t+u(b-s)c-c(b-s)u)\nonumber\\
+s(c(d-b)t-t(b-s)c+t(d-b)u-u(d-b)t+u(d-b)c-c(d-b)u)\nonumber\\
+b(c(s-d)t-t(s-d)c+t(s-d)u-u(s-d)t+u(s-d)c-c(s-d)u)]
\label{eq:fs}
\end{eqnarray}
For the KM matrix we use the standard parametrization of Ref.\cite{ok}
where the CP-odd invariant is
\be
\tilde{\delta}=\sin \delta c_1c_2c_3s_1^2s_2s_3.
\ee
The letters $u,\;d,\;s,\;c,\;b,\;t$ denote here the Green's functions of
the corresponding quarks. Each product of four quark propagators allows for
cyclic permutations of the kind
\[
udcs=dcsu=csud=sudc.
\]

{}From expression (\ref{eq:fs}) it follows in particular that any diagram
should be antisymmetrized in the masses $m_1$ and $m_3$ of the quarks
adjoint to its upper block. This upper block can be either mass or vertex
operator depending on where the photon is attached to. This antisymmetry
property is sufficient for vanishing of the first term of the expansion in
the photon momentum $q$, this term corresponding to EDM \cite{KhP}. We will
demonstrate below that the next term of the expansion in $q$ does not turn
to zero after the antisymmetrization and summation over all flavors.

{}From all the possibilities of the photon attachment presented in Figs.2 -
8, diagrams 5 - 8 only can contribute to the effect. Diagrams 2 - 4 vanish
after the antisymmetrization in the masses $m_1$ and $m_3$. Let us
determine first the spinor structure of the upper part of the quark loop
in diagrams 5 - 8, the mass or vertex operator together with both
adjoint quark propagators.  The left-handed projectors $1+\gamma_5$
present in the weak interaction vertices, single out in it vector or axial
structure only. The type of the MQM interaction with the electromagnetic
field gradient (\ref{eq:MQM}) we are looking for, fixes the general
structure of that part (up to multiplying by $\gamma_5$) as:
\be
h(p^2)(pq)\gm p_{\nu} \FD_{\mu\nu},
\label{eq:gs}
\ee
where $p$ is the momentum inside the quark loop (see Fig.5) and $h(p^2)$ is an
invariant function dependent on $p^2$ and quark masses squared. It can be
easily checked that after contracting with the $\gamma$-matrix structure
left and integrating over $p$ expression (\ref{eq:gs}) produces indeed
stucture (\ref{eq:MQM}) we are looking for.

Let us consider now in more detail how structure (\ref{eq:gs}) arises.
The complete expression for the upper part of diagrams 5 - 8 can be
written as follows:
\begin{eqnarray}
A_{\mu}(1+\gf)[e_1S_1(p-q)\gm
S_1(p)\Sigma(p)S_{3}(p)+S_1(p-q)\Gamma_{\mu}(p-q,p)S_3(p)\nonumber\\
+e_1S_1(p-q)\Sigma(p-q)S_3(p-q)\gm S_3(p)](1-\gf).
\label{eq:5-8}
\end{eqnarray}
Here $A_{\mu}$ is the electromagnetic vector-potential, $e_1$ is the
charge of the quark with mass $m_1$; $S_i(p)=(\ph-m_i)^{-1}$. The mass
operator and the vertex part are denoted as $\Sigma$ and $\Gamma_{\mu}$
respectively. The projectors $1\pm \gf$ originate from the external W-boson
interaction vertices.

The zeroth and first terms of the expansion of expression (\ref{eq:5-8}) in
$q$ vanish after the antisymmetrization in the indices 1 and 3 \cite{KhP}.
The second term of the expansion consists of several contributions. It
should be noted first of all that the contribution corresponding to the
second-order term in the expansion of $\Gamma_{\mu}$ in $q$ vanishes.
Indeed, it is symmetric in the masses $m_1$ and $m_3$ since in the propagator
$S_1$ the momentum $q$ can be put equal to zero. The contributions left
can be conveniently split into two groups. To the first one belong those
terms where the first term of the $\Gamma_{\mu}$ expansion in $q$ is
multiplied by the first term in the expansion of the propagator $S_1$. To
the second one belong all the contributions with the mass operator and
vertex part taken at the vanishing $q$.

The first group of contributions does not need renormalization and the
corresponding interim result can be obtained rather easily:
\begin{eqnarray}
A_{\mu}(1+\gf)[\fr{\ph-\qh}{(p-q)^2-m_1^2}q_{\nu}
\left.\fr{\partial{\Gamma_{\mu}}}{\partial{q_{\nu}}}\right|_{q=0}\,
\fr{\ph}{p^2-m_3^2}](1-\gf)\,-\,(m_1\leftrightarrow m_3)\nonumber\\
\longrightarrow \fr{2e_2(m_1^2-m_3^2)h(p^2)p^2(pq)\gm p_{\nu} \FD_{\mu\nu}}
{(p^2-m_1^2)^2(p^2-m_3^2)^2}(1-\gf).
\label{eq:1}
\end{eqnarray}
Here $h(p^2)$ denotes a result of the loop integration which will be
explicitly calculated below, $e_2$ is the charge of the quark inside the
vertex.

To investigate the second group of contributions we will need expressions
for the mass operator and vertex part \cite{Shab,KhP}. Unrenormalized mass
operator in the V-A theory is simple:
\be
\Sigma= \ph (1+\gf)f(p^2)
\ee
The renormalization introduces into the operator $\Sigma$ the dependence
on external masses:
\be
\Sigma=\ph (1+\gf)\tilde{f}(p^2)-f_{13}[\ph  (1-\gf)-m_1(1-\gf)-m_3(1+\gf)],
\ee
where $f_{13}$ and $\tilde{f}$ are expressed via the function $f$ and
masses $m_1,\;m_3$ as follows:
\[
\tilde{f}(p^2)=f(p^2)-\fr{m_1^2f_1-m_3^2f_3}{m_1^2-m_3^2},\;\;\;
f_{13}=\fr{m_1m_3(f_1-f_3)}{m_1^2-m_3^2};\;\;
f_i=f(p^2=m_i^2), \; \; i=1,\,3.
\]

The vertex part renormalization at the vanishing momentum of external
photon should comply with the Ward identity:
\be
\Gamma_{\mu}(p,p)=-e_1\fr{\partial \Sigma}{\partial p_{\mu}},
\label{eq:WI}
\ee
where $e_1$ is the charge of the quark with mass $m_1$. This identity allows
one to reduce the second group of contributions to
\be
e_1A_{\mu}(1+\gf)[S_1(p-q)\gm
S_1(p)\Sigma(p)S_{3}(p)-S_1(p-q)\fr{\partial \Sigma}{\partial p_{\mu}}S_3(p)
+S_1(p-q)\Sigma(p-q)S_3(p-q)\gm S_3(p)](1-\gf).
\label{eq:5-8a}
\ee
The substitution of the renormalized operator $\Sigma$ into this
expression and its expansion in $q$ leads to a simple result. The sum of
the terms proportional to $f_{13}$ is symmetric in the masses and
therefore does not contribute to the effect. A nonvanishing contribution
originates from that term in the mass operator which is proportional to
$\ph(1+\gf)$. Omitting intermediate steps, we obtain for the second group
contribution the following expression:
\be
 \fr{2(m_1^2-m_3^2)\tilde{f}'p^2(pq)\gm p_{\nu} \FD_{\mu\nu}}
{(p^2-m_1^2)^2(p^2-m_3^2)^2}(1-\gf).
\label{eq:2}
\ee
where $\tilde{f}'=\partial{\tilde{f}}/\partial{(p^2)}$. Curiously enough,
the renormalization counterterms do not contribute at all to the effect
the derivative of $\tilde{f}$ in $p^2$ coincides with that of the
unrenormalized function $f$:
\[
\fr{\partial{\tilde{f}}}{\partial{(p^2)}}=\fr{\partial{f}}{\partial{(p^2)}}.
\]

Both expressions, (\ref{eq:1}) and (\ref{eq:2}), produce contributions to the
W-boson MQM. Indeed, their contraction with the gamma-matrix
structure left allows one to write down the effective interaction of the
electromagnetic field gradient with W-boson as follows:
\be
H_{eff}=\tilde{\delta}\fr{g^4}{2}W_{\mu}^*W_{\nu}\sum_{flavour}\!\int\!\fr{d^4p}{(2\pi)^4}
 \fr{(pq)p_{\beta}(p_{\nu}\FD_{\mu\beta}+p_{\mu}\FD_{\nu\beta})
 (m_1^2-m_3^2)[e_1f'+e_2h]p^2}
{(p^2-m_1^2)^2(p^2-m_3^2)^2[(k-p)^2-m_4^2]}.
\label{eq:int}
\ee
In this expression $g$ is the semiweak charge, $m_4$ is the mass of the
quark in the lower fermion line. One can easily check that after the
integration over $p$ interaction (\ref{eq:MQM}) arises. Indeed, in the result
$(pq)p_{\beta}p_{\nu(\mu)}$ is substituted for by $q_{\nu(\mu)}k_{\beta}$
which reduces (\ref{eq:int}) to (\ref{eq:MQM}). The exact cancellation of
contributions (\ref{eq:1}) and (\ref{eq:2}) is impossible since the functions
$h(p^2)$ and $f'(p^2)$ depend differently on $p^2$ and the masses of the
quarks inside the mass operator and vertex part.

4. After convincing ourselves in the absence of the exact cancellation of
the W-boson MQM in two-loop approximation, we are going to find its value.

It is natural to consider all quark masses but $m_t$ small as compared to the
W-boson one $M$. Together with the quark mass hierarchy it allows one to
simplify the calculations considerably, restricting to those contributions to
MQM which are of lowest order in the light quark masses. Besides, it is
also natural to single out the contributions with logarithms of large mass
ratios, e.g., $\log(m_t/m_c),\,\log(m_b/m_s),\,\log(M/m_b)$ etc.

All the diagrams can be split into two types, depending on which
quarks, $U\;(u,\;c,\;t)$ or $D\;(d,\;s,\;b)$, flow inside the mass or vertex
operator. It is convenient to sum first of all over the flavours of the quarks
masses of which were denoted up to now as $m_1$ and $m_3$. For the two
types mentioned we get respectively:
\begin{eqnarray}
\sum  \fr{(m_1^2-m_3^2)}{(p^2-m_1^2)^2(p^2-m_3^2)^2}\longrightarrow
 \fr{-m_b^4m_s^2}{p^4(p^2-m_b^2)^2(p^2-m_s^2)^2} \nonumber\\
 \fr{-m_t^4m_c^2}{p^4(p^2-m_t^2)^2(p^2-m_c^2)^2}.
\label{eq:sum}
\end{eqnarray}
In expression (\ref{eq:sum}) we put $m_u=m_d=0$. We can determine now the
characteristic momenta $p$. When quarks are arranged according to
the second line of formula (\ref{eq:sum}), integral (\ref{eq:int}) is
infrared divergent if one neglects the masses $m_s$ and $m_b$ in the
denominator. It means that the typical loop momenta contributing to the
effect are $p\sim m_b$. In the opposite case when D-quarks are inside the
mass or vertex operator, the typical momenta range is large: $p\sim M$.

Now we have to sum over the flavors left. Here the difference is essential
in the dependence of $f'(p^2)$ and $h(p^2)$ on quark masses. The function
$f$ allows for the expansion in the mass for the case of a light quark
inside the mass operator:
\be
f'(p^2,m^2)=f'(p^2,m^2=0)+\left.m^2\fr{df'}{d(m^2)}\right|_{m=0}+...
\label{eq:ef}
\ee
We neglect the terms of higher order in a light quark mass. For D-quarks
inside the mass operator one can easily demonstrate the cancellation of
the terms of zeroth, second and fourth powers in mass:
\be
\fr{f'(m^2_b)-f'(m^2_s)}{(k-p)^2-m_d^2}+
\fr{f'(m^2_s)-f'(m^2_d)}{(k-p)^2-m_b^2}+
\fr{f'(m^2_d)-f'(m^2_b)}{(k-p)^2-m_s^2}={\cal O}(m_b^4m_s^2)
\label{eq:sumd})
\ee
Therefore this contribution to the MQM is ${\cal O}(m_b^4m_c^2m_s^2)$,
i.e., it is suppressed as the eighth power of the ratio of light quark
masses to that of W. In the opposite group of diagrams one cannot expand
in $m_t$, so the summation left introduces the suppression $\sim m_c^2$:
\be
 \sum \fr{f'(p^2,m^2)}{(k-p)^2 -m_4^2)}\longrightarrow
-\left.\fr{df'}{d(m^2)}\right|_{m=0}\,\fr{m_c^2}{(k-p)^2-m_t^2}+f'(m_t^2)\fr{m_c^2}{(k-p)^4}.
\ee
Since the integral over $p$ is dominated by momenta $p\sim m_b$ (see
formula (\ref{eq:sum})), the discussed contribution to the MQM is
${\cal O}(m_b^2m_c^2m_s^2)$. However, we will not to calculate it since
this contribution is small as compared with another term in expression
(\ref{eq:int}) which contains the function $h(p^2,m^2)$.

The point is that, as distinct from the function $f,\;h$ does not allow
for a simple expansion in a light quark mass. For a light quark inside the
vertex part, $h$ contains terms proportional to $m^2\log m^2$. Therefore,
the terms containing $m_s^2m_b^2$ in the expression analogous to
(\ref{eq:sumd}), not only do not cancel, but are enhanced as
$\log(m_b^2/m_s^2)$. Let us demonstrate this assertion in more detail. To
this end we write down the vertex integral explicitly (in fact its dominating
part
where the quark interacts with the external field) and sum over
the masses of all quarks in the upper and lower propagators of the
quark loop. The result is:
\be
\fr{g^2e_2m_s^2m_b^6}{((k-p)^2-m_s^2)((k-p)^2-m_b^2)}\int\fr{d^4l}{(2\pi)^4}
\fr{\FD_{\mu\nu}(\gm l_{\nu}+\fr{2}{M^2}\hat{l}l_{\mu}p_{\nu})(1+\gf)}
{l^2(l^2-m_s^2)^2(l^2-m_b^2)^2[(p-l)^2-M^2]}
\label{eq:vp}
\ee
It is obvious now that the integral is dominated by the momenta region
$m_s\ll l \ll m_b$ and contains a large logarithm. Expanding the
denominator in the ratio $(pl)/(p^2-M^2)$ we obtain
\be
\fr{g^2e_2m_s^2m_b^2\log(m_b^2/m_s^2)}{16\pi^2((k-p)^2-m_s^2)((k-p)^2-m_b^2)}
\left[ \fr{1}{(p^2-M^2)^2}+\fr{1}{p^2-M^2}\right]
\FD_{\mu\nu}\gm p_{\nu}(1+\gf).
\label{eq:log1}
\ee
This formula is valid for all $p^2$ as long as $p^2-M^2\gg Mm_b$.

The last integral over $p$ contains also a large logarithm when
W-boson is on mass shell. It can be easily checked indeed that the
integral diverges logarithmically if one neglects quark masses in the
denominator of the factor preceding the integral in formula (\ref{eq:vp})
and put $k^2=M^2$. Retaining in the last integral that contribution only
which is enhanced by a large logarithmic factor $\log(M^2/m_b^2)$, we
obtain the final expression for the CP-odd interaction of W-boson with
the electromagnetic field gradient:
\be
H_{eff}=\tilde{\delta}\fr{eg^4}{12(16\pi^2)^2}\fr{m_s^2m_c^2m_b^2}{M^4}\fr{m_t^4}
{(m_t^2-M^2)^2}
\log(m_b^2/m_s^2)\log(M^2/m_b^2)W_{\mu}^*W_{\nu}(\partial_{\mu}\FD_{\nu\alpha}+
\partial_{\nu}\FD_{\mu\alpha})k_{\alpha}.
\label{eq:otvet}
\ee

As to another group of diagrams, with $U$-quarks in the vertex part, they
also contain $\log[(M^2-p^2)/m_c^2]$ in a formula analogous to
(\ref{eq:log1}), but cannot compete with (\ref{eq:otvet}) since they have
no log in the integral over $p^2$.

Going over to the Fermi constant $G_F=\sqrt{2}g^2/(8M^2)$ we get the final
formula for the W-boson MQM in the Kobayashi-Maskawa model:
\be
{\cal
M}=-\fr{e}{48\pi^4}\tilde{\delta}G_F^2\fr{m_s^2m_c^2m_b^2}{M^4}\fr{m_t^4}{(m_t^2-M^2)^2}
\log(m_b^2/m_s^2)\log(M^2/m_b^2).
\ee

The extreme smallness of the MQM in the standard model is in no way
unexpected. In other models of CP-violation ${\cal M}$ can well turn out
much larger.

We are greatly indebted to C.P. Burgess, C. Hamzaoui, G. Couture,
G. Jakimow and I. Maksymyk for stimulating discussion and correspondence.

\newpage
\setlength{\unitlength}{1mm}

\begin{figure}
\begin{picture}(200,200)(0,0)
\put(10,60){\oval(15,30)}
\hpl{3}{68}{1.5}{3}{5}
\hpl{-6}{52}{1.5}{3}{3}
\hpl{18}{52}{1.5}{3}{3}

\multiput(3,60)(-2.5,0){3}{\line(-1,0){1.5}}
\multiput(168,60)(2.5,0){3}{\line(1,0){1.5}}
\multiput(60,75)(0,2.5){3}{\line(0,1){1.5}}
\multiput(110,68)(0,-2.5){3}{\line(0,-1){1.5}}
\multiput(-0.5,112)(0,-2.5){3}{\line(0,-1){1.5}}
\multiput(80,105)(0,-2.5){3}{\line(0,-1){1.5}}
\multiput(160,112)(0,-2.5){3}{\line(0,-1){1.5}}

\floop{50}{0}
\floop{100}{0}
\floop{150}{0}
\floop{0}{60}
\floop{70}{60}
\floop{140}{60}
\floop{0}{120}

\put (10,155){\makebox(0,0){Fig. 1}}
\put (10,95){\makebox(0,0){Fig. 2}}
\put (80,95){\makebox(0,0){Fig. 3}}
\put (150,95){\makebox(0,0){Fig. 4}}
\put (10,35){\makebox(0,0){Fig. 5}}
\put (60,35){\makebox(0,0){Fig. 6}}
\put (110,35){\makebox(0,0){Fig. 7}}
\put (160,35){\makebox(0,0){Fig. 8}}

\put (10,43){\makebox(0,0){k-p}}
\put (6.5,56){\makebox(0,0){p-q}}
\put (20,60){\makebox(0,0){p}}
\put (-3,48){\makebox(0,0){k-q}}
\put (23,48){\makebox(0,0){k}}

\end{picture}

\end{figure}

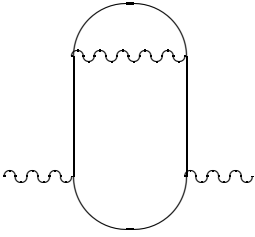
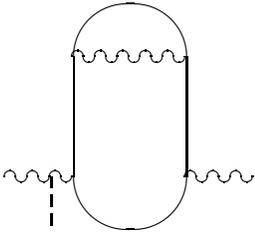
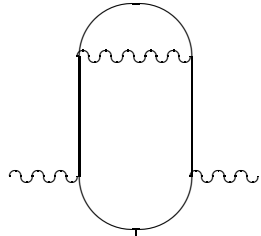
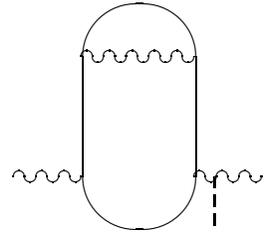
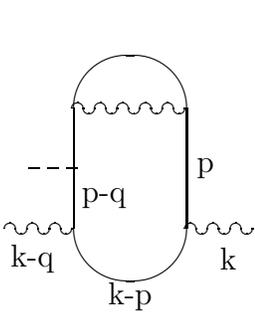
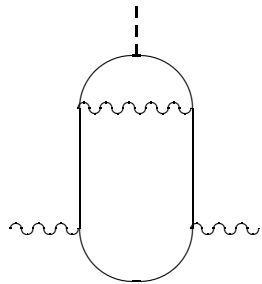
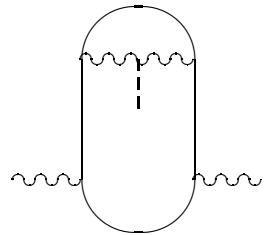
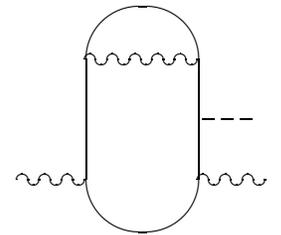
\end{document}